\documentclass[lettersize,journal]{IEEEtran}
\usepackage{amsmath,amsfonts}
\usepackage{algorithmic}
\usepackage{algorithm}
\usepackage{array}
\usepackage[caption=false,font=normalsize,labelfont=sf,textfont=sf]{subfig}
\usepackage{textcomp}
\usepackage{stfloats}
\usepackage{url}
\usepackage{verbatim}
\usepackage{graphicx}
\usepackage{cite}
\usepackage{booktabs}
\usepackage{hyperref}
\usepackage{graphicx}
\usepackage{dcolumn}
\usepackage{bm}
\usepackage{subcaption}
\usepackage{xcolor}
\usepackage{setspace}

\begin{document}

\title{Modeling electrothermal feedback of superconducting nanowire single photon detectors in SPICE}

\author{Hanson Nguyen, Alejandro Simon\textsuperscript{*}, Reed Foster, and Karl~K. Berggren,~\IEEEmembership{Fellow,~IEEE}%
\thanks{This work was supported by the U.S. National Science Foundation under Grant EEC-1941583.}%
\thanks{Hanson Nguyen is with the School of Electrical, Computer and Energy Engineering, Arizona State University, Tempe, AZ 85287 USA, and also with the Department of Electrical Engineering and Computer Science, Massachusetts Institute of Technology, Cambridge, MA 02139 USA (e-mail: hansonng@mit.edu).}%
\thanks{Alejandro Simon, Reed Foster, and Karl~K. Berggren are with the Department of Electrical Engineering and Computer Science, Massachusetts Institute of Technology, Cambridge, MA 02139 USA (e-mail: alejansi@mit.edu; reedf@mit.edu; berggren@mit.edu).}
\thanks{\textsuperscript{*}Corresponding author: alejansi@mit.edu}}



\date{\today}

\maketitle

\begin{abstract}
Superconducting nanowire single-photon detectors (SNSPDs) exhibit complex switching behaviors due to electrothermal feedback during the detection process. Modeling and understanding these behaviors is integral for designing superconducting devices; however, many models often prioritize accuracy over computational speed and intuitive integration for circuit designers. Here, we build upon a growing architecture of SPICE tools for superconducting nanowire devices by capturing complex residual heating effects in a compact thermal model of an SNSPD. We demonstrate that our model is comparable to more complicated thermal models of superconducting nanowire devices, including finite-element simulations, and is applicable for the fast development of SNSPD circuits.
\end{abstract}

\begin{IEEEkeywords}
Superconducting integrated circuits, electrothermal modeling, superconducting nanowire single-photon detectors, circuit simulation, SPICE
\end{IEEEkeywords}

\section{\label{sec:Introduction}Introduction}

Superconducting nanowire single-photon detectors (SNSPDs) are becoming increasingly popular in areas such as quantum optics, fiber optic communications, and LiDAR systems due to their record low jitter times of $\mathrm{sub-}3 \, \mathrm{ps}$ \cite{Korzh2020}, ultra-low dark count rates of $10^{-4} \, \mathrm{Hz}$ \cite{Shibata2015}, and high single-photon detection efficiencies of $\sim 98\%$ \cite{Reddy2020}. Such developments have led to the widespread adoption of SNSPDs in applications such as integrated nanophotonics\cite{Zhu2020, Zhu2021}, single-flux quantum read-out arrays\cite{Terai2012}, and time-correlated single-photon counting circuits\cite{Shcheslavskiy2016}. However, optimizing device performance in these systems requires accurate modeling of electrothermal effects, e.g., substrate-mediated heating. These effects are central to detector performance, impacting key metrics like detection efficiency, reset time, and dark count rates, yet they are poorly understood. A compact, physics-based model that captures these interdependent effects in a schematic representation would enable the reliable integration and design of SNSPDs into advanced circuits.

Through the industry-standard Simulation Program with Integrated Circuit Emphasis (SPICE), compact device models of various superconducting nanowire-based devices have been designed with accurate timing and signal amplitude properties \cite{Berggren2018, Castellani_2020a}. Such models have enabled superconducting nanowire circuits for neuromorphic computing\cite{Toomey2019}, binary shift registers \cite{Foster2023}, and multi-pixel SNSPD arrays \cite{Hao2024}, adding to the growing family of superconducting nanowire electronics \cite{medeiros2025scalablesuperconductingnanowirememory, Buzzi2023, Castellani2024, Butters2021, McCaughan2019}. Although their simplicity enables fast simulation compared to finite-element methods, current SNSPD SPICE models overlook electrothermal effects that cause complex switching dynamics. For example, well-known effects such as after-pulsing are not captured, and phenomenological parameters restrict the physical insight gained from an SNSPD model. Thus, implementing electrothermal dynamics after photon detection into an SNSPD SPICE model will enable circuit designers to predict undesirable behaviors for larger SNSPD circuits. Here, we develop a macroscopic, thermal-based SNSPD model via a lumped electrothermal circuit model implemented in SPICE. We demonstrate that this model can capture thermal effects such as after-pulsing, hotspot growth and stabilization, and latching while still allowing for fast integration into superconducting nanowire circuits. 

\IEEEpubidadjcol

\begin{figure}[t!]
    \centering
    \includegraphics[width=0.70\columnwidth]{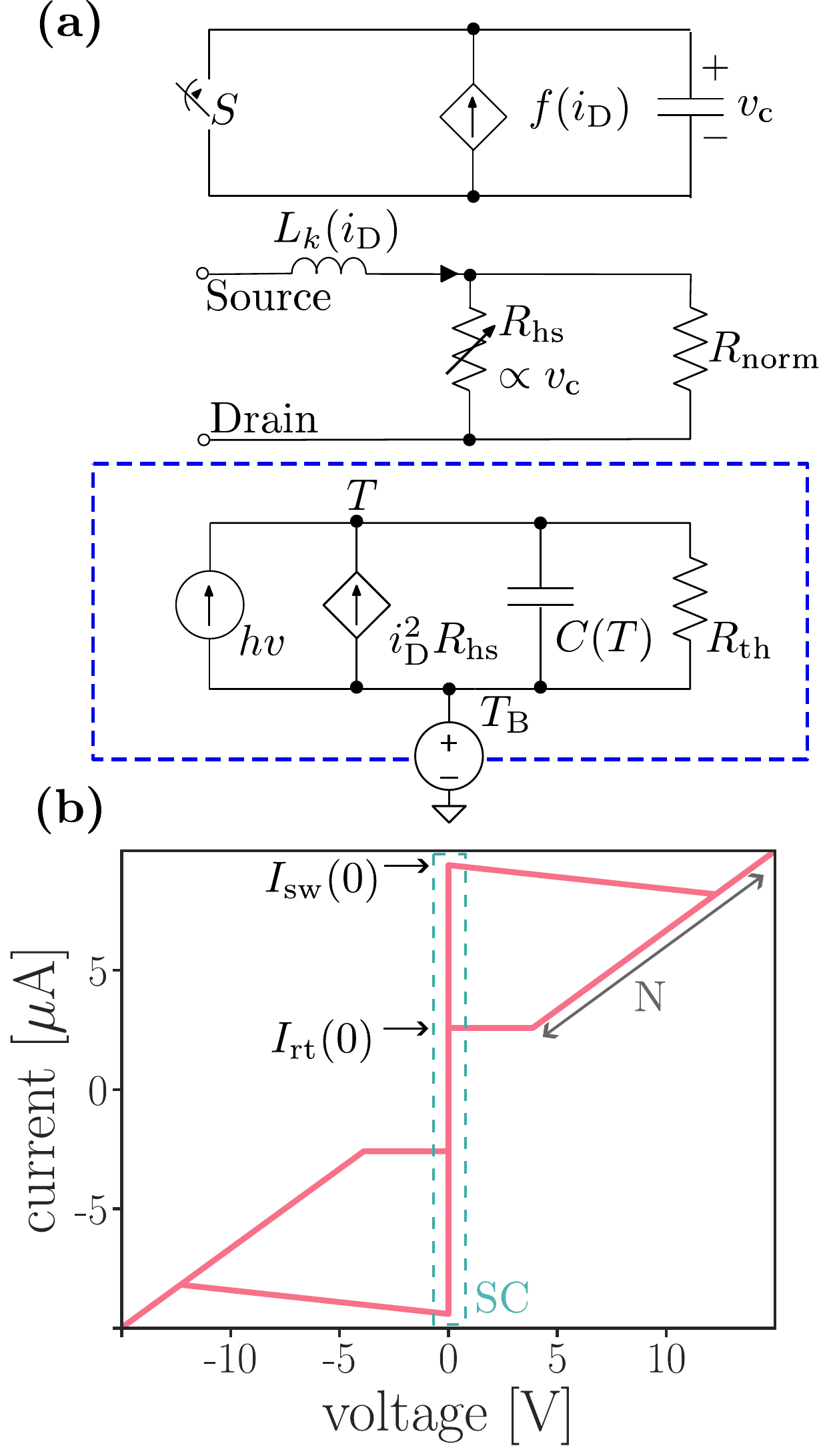}
    \caption{(a) SPICE SNSPD model from Ref.~\cite{Berggren2018}, showing hotspot growth $f(i_\mathrm{D})$ and added electrothermal feedback (blue dashed RC network mapping heat capacitance and thermal resistance to $C(T)$ and $R_{\rm th}$), with a pulsed source $hv$ and Joule heating source $i_\mathrm{D}^2R_{\rm hs}$. (b) Simulated $I$–$V$ curve marking the zero-temperature retrapping current $I_{\rm rt}(0)$, switching current $I_{\rm sw}(0)$, and the superconducting (SC) vs.\ normal (N) regions.}
    \label{fig:Circuit Model}
\end{figure}

\section{\label{sec:Methods}Methods} 

Fig.~\ref{fig:Circuit Model}(a) shows the SPICE model, which incorporates the thermal-boundary velocity model from Ref.~\cite{Berggren2018} for the hotspot growth and our electrothermal feedback model for the wire temperature \cite{Kerman2006, Kerman2009}. The switching current is the threshold current at which the nanowire switches from the superconducting state to the normal state, dictating whether the hotspot resistor $R_\mathrm{hs}$ is zero or nonzero. The switching current is a function of the temperature $T$, described by the expression
\begin{equation}\label{eq:phenomenological switching current}
I_\mathrm{sw}(T) = I_\mathrm{sw}(0)\Big(1-\Big(\frac{T}{T_\mathrm{c}}\Big)^2\Big)^{3/2},
\end{equation}
where ${I_\mathrm{sw}(0)}$ is the zero temperature switching current \cite{Yang2007}. We can model the temperature of the nanowire with the one-dimensional time-dependent heat equation
\begin{equation}\label{eq:heat equation}
J^2\rho + \kappa\frac{\partial^2 T}{\partial x^2} - \frac{h_c}{d}(T^n-T_\mathrm{B}^n) = c\frac{\partial T}{\partial t},
\end{equation}
where $J$ is current density, $\rho$ is resistivity, $T_\mathrm{B}$ is the bath temperature, $h_c$ is the thermal conductivity, $d$ is the thickness, $\kappa$ is the thermal diffusivity, and $c$ is the heat capacity. In our model, we use the linear approximation $n = 1$, but we note that the fitting parameter $n$ can be adjusted to capture various effects that are involved in the hotspot cooling process \cite{Dane2022}.

Since cooling along the nanowire length is significant only when the hotspot is small, we simplify the spatial dependence by replacing the cooling term, $\kappa\frac{\partial^2 T}{\partial x^2}$, with a phenomenological scaling factor $K$. In doing so, we assume the nanowire temperature is uniform and that $\kappa$ does not vary strongly with position or temperature \cite{Berggren2018}. Thus, Eq. \eqref{eq:heat equation} becomes
\begin{equation}\label{eq:reduced heat equation}
    \frac{J^2\rho}{K} - \frac{h_c}{d}(T-T_\mathrm{B}) = c\frac{\partial T}{\partial t}.
\end{equation}
This time-dependent thermal equation can be implemented in SPICE with a parallel RC circuit  
\begin{equation}\label{eq:RC circuit equation}
    I - \frac{v_c}{R} = C\frac{dv_c}{dt}.
\end{equation}
In the RC thermal network, we use $I = \frac{J^2\rho}{K} V = i_\mathrm{D}^2R_\mathrm{hs}$, $v_c = T-T_\mathrm{B}$, $R = \frac{d}{h_c}v$, and $C = cV$, where $V$ is the volume of the normal region. We then implement the temperature dependence of both the normal and superconducting state heat capacity. The heat capacity of the nanowire in the normal state is given by the Einstein-Debye model
\begin{equation}\label{eq:nonlinear heat capacity}
    c_\mathrm{n}(T) = c_\mathrm{el}(T) + c_\mathrm{ph}(T) =\gamma T + \alpha T^3,
\end{equation}
where $c_\mathrm{el}(T)$ and $c_\mathrm{ph}(T)$ are the electronic and phononic contributions to the heat capacitance and $\gamma$ and $\alpha$ are material-dependent constants. We can explicitly write $\gamma$ and $\alpha$ in the low-temperature approximation as 
\begin{align}
    \gamma &= \frac{\pi^2nk_\mathrm{B}^2}{2E_\mathrm{F}} \\
    \alpha &= \frac{12\pi^4N_\mathrm{i}k_\mathrm{B}}{5T_\mathrm{D}^3}
\end{align} 
where $n$ is the free electron density, $N_\mathrm{i}$ is the ion density, $k_\mathrm{B}$ is the Boltzmann constant, $E_\mathrm{F}$ is the Fermi energy level of the material, and $T_\mathrm{D}$ is the Debye temperature. For our results, we use $\gamma = 64.7 \,\mathrm{J/m^3\,K^2}$ and $\alpha = 2.38 \,\mathrm{J/m^3\,K^4}$, which is consistent with NbN \cite{Chockalingam2008}. Changing these material parameters alters the nonlinear heat capacity of the RC thermal network, which in turn affects the characteristic thermal relaxation time $\tau_\mathrm{th}$. 

\begin{figure}[t!]
     \centering
    \includegraphics[width=\columnwidth]{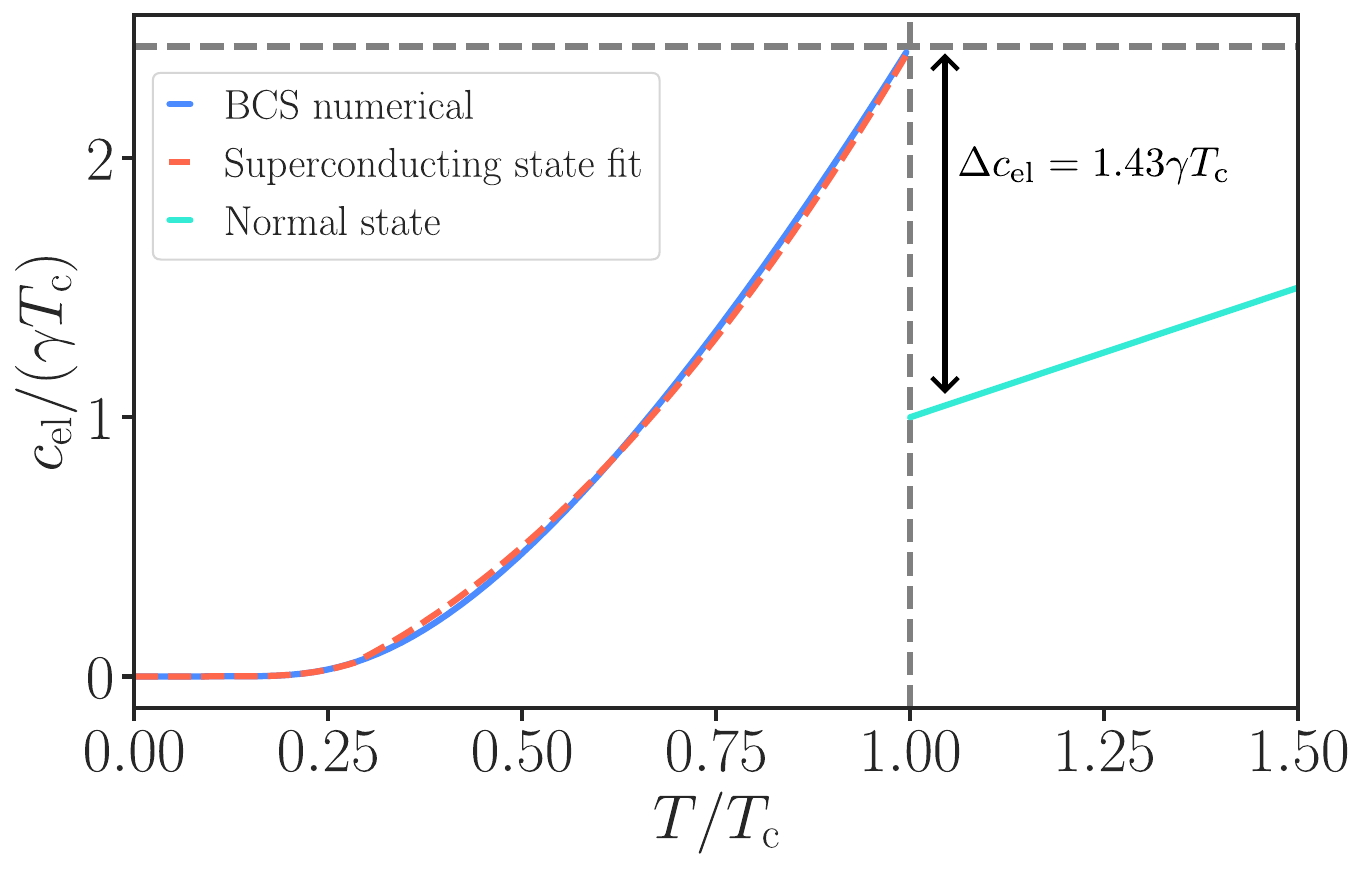}
     \caption{The electronic heat capacity fitting function used in the model compared to the numerically solved BCS heat capacity. The plot shows the BCS characteristic jump in electronic heat capacity at $T = T_\mathrm{c}$ such that $\Delta c_\mathrm{el}~=~c_\mathrm{el,sc}(T_\mathrm{c})~-~c_\mathrm{el,n}(T_\mathrm{c})~=~1.43\gamma T_\mathrm{c}$.}
     \label{fig:Heat capacitance fitted plot}
 \end{figure}

In the superconducting state, the electronic contribution to heat capacity differs. Typically, the exact heat capacity can be solved through numerical methods. To work within the framework of SPICE, we can avoid extensive computations by using a fitted function of the form $c_\mathrm{sc}(T) = \gamma T_\mathrm{c}\alpha e^{-\beta T/T_\mathrm{c}}$ for $T<0.272T_\mathrm{c}$ and a quadratic fit of the form $c_\mathrm{sc}(T) = a+b\frac{T}{T_\mathrm{c}}+c(\frac{T}{T_\mathrm{c}})^2$ for $T>0.272T_\mathrm{c}$. We used $\alpha=8.5$, $\beta=1.44$, $a = -0.113$, $b = -0.183$, and $c = -2.775$ \cite{Johnston2013}. We note that this model may change for materials where the ideal weak-coupling BCS behavior does not apply. The resultant heat capacity in the superconducting state becomes
\begin{equation}\label{eq:superconducting_heat_capacity}
c_{\mathrm{sc}}(T)=
\begin{cases}
\gamma T_\mathrm{c}\,8.5\,e^{-1.44T/T_\mathrm{c}} + c_\mathrm{ph}(T), & T \leq 0.272\,T_\mathrm{c},\\
\gamma T_\mathrm{c}\Biggl(-0.113 - 0.183\,\dfrac{T}{T_\mathrm{c}} - \\
\quad\quad\quad 2.775\,\Bigl(\dfrac{T}{T_\mathrm{c}}\Bigr)^2\Biggr) + c_\mathrm{ph}(T), & T > 0.272\,T_\mathrm{c}.
\end{cases}
\end{equation}
In Fig.~\ref{fig:Heat capacitance fitted plot}, we show our fitted function, which has strong agreement with the numerically solved BCS heat capacity. Eqs.~\eqref{eq:phenomenological switching current},~\eqref{eq:heat equation},~\eqref{eq:nonlinear heat capacity},~and~\eqref{eq:superconducting_heat_capacity} are used to obtain the temperature in our compact model.

The bath temperature $T_\mathrm{B}$ is implemented as a DC voltage source as seen in Fig.~\ref{fig:Circuit Model}. To model the nonlinearity and discontinuity in the nanowire's heat capacity, we model the heat flow as two current sources:
\begin{align}\label{eq:heat flow as current}
I_\mathrm{n} &= c_\mathrm{n}(T)\frac{\mathrm{d}V}{\mathrm{d}t} \\ I_\mathrm{sc} &= c_\mathrm{sc}(T)\frac{\mathrm{d}V}{\mathrm{d}t}
\end{align}
for the normal state and superconducting state, with the difference being the nonlinear capacitance model used in each expression. 

  We model photon absorption as a pulsed current source in the RC thermal network, with the area of the pulse equal to the energy of the photon. This calculation of the photon energy transferred into thermal energy neglects the more intricate non-equilibrium effects for photon detection, which have previously been modeled using an \textit{ab initio} approach by integrating kinetic equations and density functional theory \cite{Simon2025}. To capture these effects, a fitting parameter is included to adjust the energy of the photon.

Fig.~\ref{fig:Circuit Model}(b) shows a simulated \textit{i-v} curve with a $10\, \mathrm{M}\Omega$ load resistor and variable current source in parallel with our SNSPD model. The SNSPD has a switching current of $9.4 \,\mu\mathrm{A}$. The current source is swept from $0 \,\    \mu\mathrm{A}$ to $12 \,\mu\mathrm{A}$, then to $-12 \,\mu\mathrm{A}$, and back to $0 \,\mu\mathrm{A}$. Initially, the nanowire is superconducting. When $|i_\mathrm{D}|$ exceeds $I_\mathrm{sw}(T)$, the nanowire transitions to the normal state, behaving as an ohmic resistor. As the current decreases toward $0 \,\mu\mathrm{A}$, $|i_\mathrm{D}|$ falls below the retrapping current $I_\mathrm{rt}(T)$, causing the hotspot to gradually shrink until superconductivity is fully restored. To better capture the dynamics of retrapping, $I_\mathrm{rt}(T)$ is temperature dependent, defined as
\begin{equation}
    I_\mathrm{rt}(T) = \sqrt{2/\psi}I_\mathrm{sw}(T),
\end{equation}
where $\psi$ is the Stekly parameter \cite{Gurevich1987}. For the hotspot growth, we used the phenomenological model of the hotspot size velocity as a function of the current through the device introduced in \cite{Kerman2009}.

\begin{figure}[h!]
    \centering
    \includegraphics[width=0.9\columnwidth]{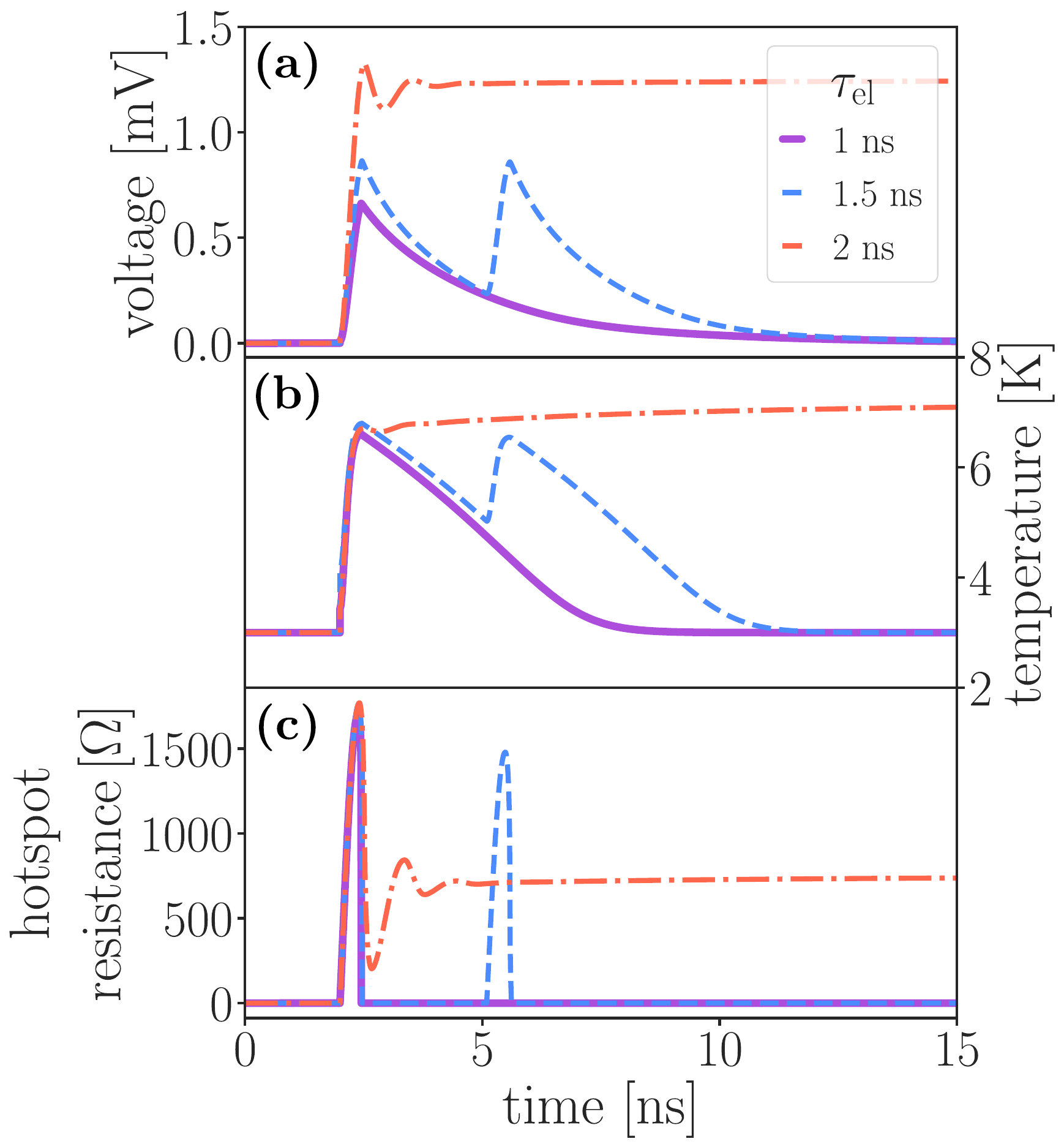}
    \caption{Response to a photon pulse at time $t = 2 \, \mathrm{ns}$ for various electronic reset times  $\tau_\mathrm{e} =2.0\,\mathrm{ns}$ (single pulse), $1.5 \,\mathrm{ns}$ (after-pulse), and $2.0\,\mathrm{ns}$ (latching). The thermal reset time is kept constant, so increasing the electronic reset time determines the device effects. (a) shows the output voltage, (b) shows the device temperature, and (c) shows the hotspot resistance. }
    \label{fig:stacked_readout}
    \end{figure}

\section{\label{sec:Results}Results}
By changing the load resistance connected to an SNSPD, we can observe the interplay between the electronic reset time $\tau_\mathrm{e}$ and $\tau_\mathrm{th}$ that results in various device effects. Fig.~\ref{fig:stacked_readout} demonstrates various effects due to different load resistances in parallel with the nanowire that control $\tau_\mathrm{e}$. For this device, we set the width to $100\,\mathrm{nm}$ and thickness to $4\,\mathrm{nm}$. The zero-current kinetic inductance of the nanowire is kept at $L_0 = 200 \, \mathrm{nH}$ and the nanowire is biased at $I_\mathrm{B} = 7.5\,\mathrm{\mu A}$ for $T_\mathrm{B} = 3\,\mathrm{K}$, $T_\mathrm{c} = 10\,\mathrm{K}$ and $I_\mathrm{sw}(0) = 9 \, \mathrm{\mu}A$. For $R_\mathrm{load} = 100 \,\mathrm{\Omega} \, (\tau_\mathrm{e} = 2.0 \, \mathrm{ns})$, current is shunted away from the nanowire before the hotspot has grown significantly, and the resulting reduced current through the nanowire allows for rapid cooling and recovery. For $R_\mathrm{load} = 130 \,\mathrm{\Omega}  \, (\tau_\mathrm{e} = 1.5 \, \mathrm{ns})$, after-pulsing occurs when the wire returns to the superconducting state prematurely at an elevated temperature. The residual heat in the wire suppresses the switching current, causing the bias current to surpass the suppressed switching current and triggering another detection event without a photon.  For $R_\mathrm{load} = 200 \,\mathrm{\Omega}  \, (\tau_\mathrm{e} = 1.0 \, \mathrm{ns})$, the larger load resistance reduces the current diversion from the nanowire, and $i_\mathrm{D}$ remains above $I_\mathrm{sw}(T)$, resulting in latching.

\begin{figure}[h!]
    \centering
    \includegraphics[width=0.75\columnwidth]{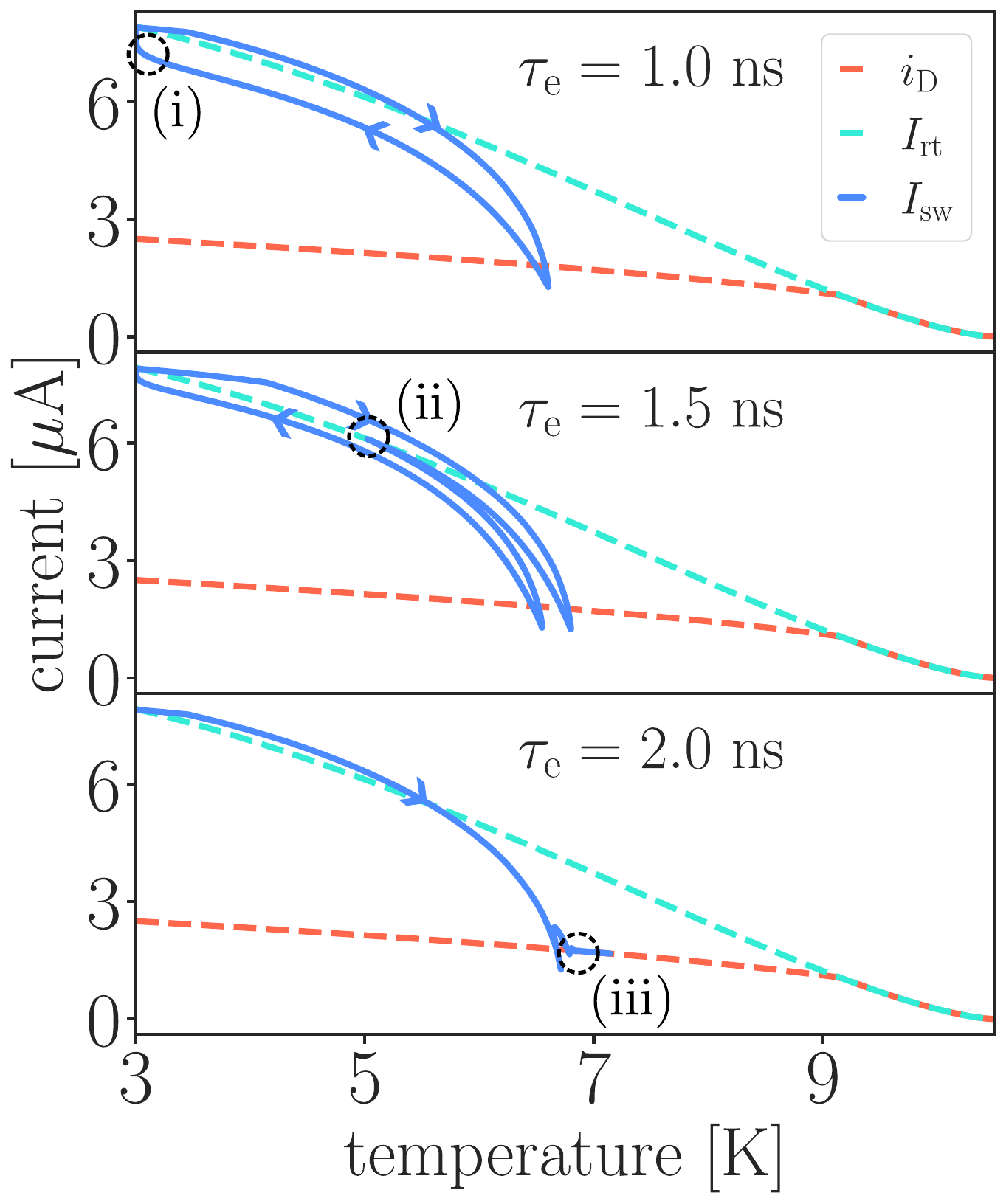}
    \caption{The current-temperature relationship for a single pulse, after-pulsing, and latching is shown. For $\tau_\mathrm{e} = 1.0 \,$ns, the nanowire cools down and the full bias current returns at the bath temperature (i). For $\tau_\mathrm{e} = 1.5 \,$ns, the nanowire cools down, but switches once (ii) before retrapping at the bath temperature. For $\tau_\mathrm{e} = 2.0 \,$ns, the nanowire does not retrap and remains at the retrapping current (iii).} 
    \label{fig:stacked_temp}
\end{figure}
False detection events and latching mechanisms can be further understood through the current-temperature relationship post-photon pulse in the same circuit configuration as described in Fig.~\ref{fig:stacked_readout}. The electronic reset process of the nanowire begins when the nanowire returns to the superconducting state and is determined by $\tau_\mathrm{e} = L_k/ R_\mathrm{load}$, where $L_k$ is the kinetic inductance \cite{Clem2012}. The thermal reset process occurs when $|i_\mathrm{D}| <I_\mathrm{rt}$ and is characterized by $\tau_\mathrm{th} = RC$.
In Fig.~\ref{fig:stacked_temp}(a), the smaller $\tau_\mathrm{e} > \tau_\mathrm{th}$ results in a single pulse. The bath temperature is $T_\mathrm{B}$ and thus $T =T_\mathrm{B}$ when $I_\mathrm{B}$ is fully restored. For $\tau_e \sim \tau_\mathrm{th}$ (Fig.~\ref{fig:stacked_temp}(b)),  $T > T_\mathrm{B}$ when $I_\mathrm{B}$ is fully restored. In this case, $I_\mathrm{B} > I_\mathrm{sw}(T)$, resulting in a false detection event. For $\tau_e < \tau_\mathrm{th}$ (Fig.~\ref{fig:stacked_temp}(c)), additional Joule heating from the larger hotspot resistance exceeds the cooling in the nanowire and sustains the resistive state, leading to a latched state. The current through the nanowire latches to the retrapping current, as $\frac{\mathrm{d}R_\mathrm{hs}}{\mathrm{d}t}$ goes to $0$.

To quantify the efficacy of our approach, we compare our model to previous electrothermal models in SPICE. Although the  heater-nanocryotron (hTron) and SNSPD operate on different principles, our primary goal is to compare the simulation speed of their SPICE models as the underlying electrothermal physics is the same. We compare the speed of our model in response to a 1550 nm detection event to a comparable current pulse in the heater of an hTron from Ref. \cite{karam2025parameterextractionsuperconductingthermal}. As suggested in Ref. \cite{ElDandachi2023Efficient}, a relative tolerance of $10^{-6}$ is chosen. We sweep the time resolutions $\Delta t_\mathrm{max}$ by increments of $1\,\mathrm{ps}$ from $\Delta t_\mathrm{max} = 1\,\mathrm{ps}$ to $\Delta t_\mathrm{max} = 50\,\mathrm{ps}$ and simulate for $100$~ns. The thermal reset time was set to $\tau_\mathrm{th} \approx 8$~ns for all simulations. The simulations were run on a $2.6$~GHz processor. The behavioral hTron model developed in Ref.~\cite{karam2025parameterextractionsuperconductingthermal} completed in $\sim11\,$s, and SNSPD model took $\sim20\,$s to complete; however, notably the SNSPD model uses significantly less fitting parameters. The marginally slower simulation speed of the SNSPD model can be explained by the stronger nonlinear stiffness due to the sharp changes in the nonlinear heat capacitance. Nonetheless, both models show improvement on many orders of magnitude compared to the finite-element models of superconducting nanowires developed in Refs.~\cite{Baghdadi2020, 10814066}.

\begin{figure}[h!]
    \centering
    \includegraphics[width=\columnwidth]{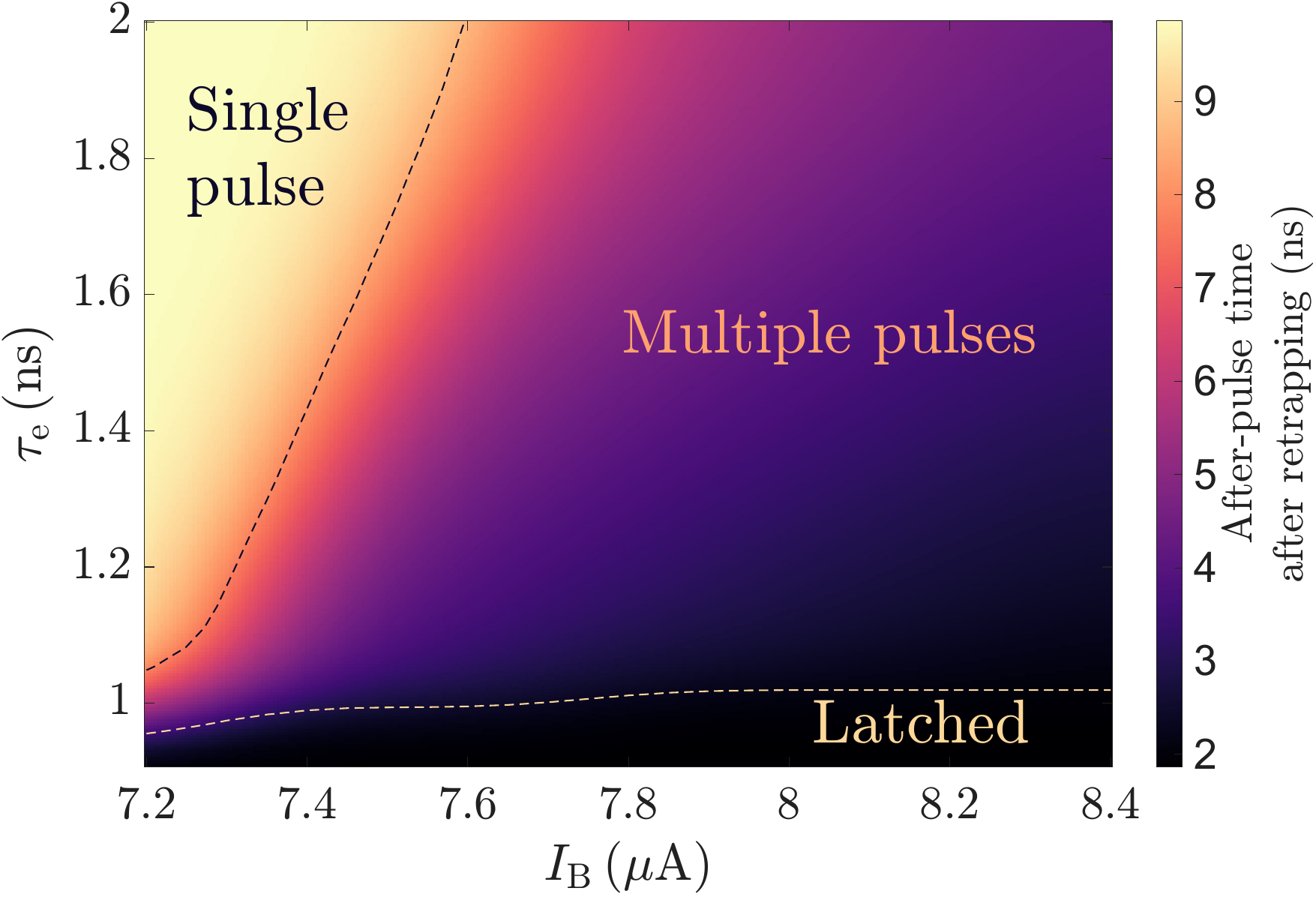}
     \caption{After-pulsing time by varying $\tau_\mathrm{e}$ and $I_\mathrm{B}$ generated by the electrothermal SPICE model. The contour lines distinguishes single pulses, after-pulsing, and latching.}
    \label{fig:I_B and R_L for after-pulsing}
\end{figure}

Finally, we demonstrate that our model can predict the boundaries between single pulses, after-pulsing, and latching in Fig.~\ref{fig:I_B and R_L for after-pulsing} by varying the bias current $I_\mathrm{B}$ and the characteristic electronic reset time $\tau_\mathrm{e}$ of an SNSPD. The contour line that distinguishes a single pulse from an after-pulse can be analytically determined through a conditional expression with an exponential fit for current of the form \begin{equation}
    i_\mathrm{D} = I_\mathrm{B} - (I_\mathrm{B}-I_\mathrm{R})e^{-t/(p\tau_\mathrm{e})},   
\end{equation}
where $I_\mathrm{R}$ is the retrapping current and $p$ is a fitting parameter that accounts for the nonlinearities of the kinetic inductance, and a fitting equation of the form \begin{equation}
    T^2 = T_\mathrm{R}^2 - qt  , 
\end{equation}
where $T_\mathrm{R}$ is the retrapping temperature and $q$ is also a fitting parameter. By assuming that the difference between the switching current $I_\mathrm{sw}(T)$ and the current in the wire $i_\mathrm{D}$ initially decreases after resetting, we can use the expression 
\begin{equation}
F(I_\mathrm{B}, \tau_\mathrm{e})= I_\mathrm{0}\Big(1 - \Big(\frac{T_\mathrm{R}^2 - q t^*}{T_\mathrm{C}^2}\Big)^{3/2}\Big)- \Big(I_\mathrm{B} - (I_\mathrm{B} - I_\mathrm{R}) e^{-t^*/(p\tau_\mathrm{e}}) \Big),
\end{equation}
such that $\frac{dF(t)}{dt}|_{t =t^*} = 0$, to find whether the $i_\mathrm{D}$ surpasses $I_\mathrm{sw}(T)$ as the wire resets and cools. Solving for $t^*$, we obtain 
\begin{align}
t^* = \frac{p\tau_\mathrm{e}}{2} \, W_0\Bigg(& \frac{8}{9} \frac{(I_\mathrm{B} - I_\mathrm{R})^2}{\left( \frac{q}{T_\mathrm{C}^2} \right)^3 I_\mathrm{C}^2 (p\tau_\mathrm{e})^3} \notag \\
& \exp\Bigg[\frac{2}{p\tau_\mathrm{e}} \left( \frac{T_\mathrm{C}^2 - T_\mathrm{R}^2}{q} \right) \Bigg] \Bigg) 
- \left( \frac{T_\mathrm{C}^2 - T_\mathrm{R}^2}{q} \right),
\end{align}
where $W_0(z)$ is the principle branch of the Lambert $W$ function. Thus, after-pulsing or latching occurs when $F(I_\mathrm{B}, \tau_\mathrm{e})~\leq~0$. We note that the fitting parameters of the equation are dependent on the geometry and material of the nanowire. Thus, our model shows qualitative agreement with the electrothermal framework of Ref.~\cite{Kerman2009}, predicting that stable operating regime of an SNSPD occurs at lower bias currents and larger $\tau_\mathrm{e}$.
   
\section{\label{sec:Conclusion}Conclusion}
Our model provides a foundation to implement more complex thermal dynamics into SPICE simulations of superconducting nanowire-based devices. The ancillary thermal circuit added to the model allows for accurate simulations of common device effects that arise due to thermal dynamics in the nanowire. We are able to control device reset through both the external readout circuit of the SNSPD and material parameters, reproducing key effects such as after-pulsing and latching. These effects are particularly relevant in modular designs such as superconducting nanowire avalanche photodetectors (SNAPs), where thermal dynamics can cause significant performance degradation \cite{Marsili2012}. Our framework also enables the modeling of device architectures such as thermally coupled SNSPDs \cite{Allmaras2020}. While the compact nature of our model precludes capturing the full device physics, such as spatially dependent thermal dynamics considered in more detailed approaches \cite{Yang2007}, simplified $0$D electrothermal device models have shown good agreement with finite-element methods, including the behavioral hTron model from Ref.~\cite{karam2025parameterextractionsuperconductingthermal}, and provide valuable insight into larger-scale device performance.

\section{\label{sec:Acknowledgements}Acknowledgments}

The authors thank Gian Luca Dolso and Phillip D. Keathley for their feedback during the preparation of this manuscript. Financial support was provided by the U.S. National Science Foundation, Grant Number EEC-1941583. HN acknowledges support from the MIT Summer Research Program. AS acknowledges support from the NSF GRFP. RF acknowledges support from the Alan McWhorter fellowship. 

\section{\label{sec:Data availibility}Data availability}

The data and model that support the findings of this study are publicly available via Zenodo at \href{https://doi.org/10.5281/zenodo.17246356}{https://doi.org/10.5281/zenodo.17246356}\cite{https://doi.org/10.5281/zenodo.17246356}.

\bibliographystyle{IEEEtran}
\bibliography{Bib}

\end{document}